\documentclass[12pt]{article}
\usepackage{amssymb,epsfig}

\renewcommand{\thefootnote}{\fnsymbol{footnote}}

\newcommand{\be}{\begin{equation}}
\newcommand{\ee}{\end{equation}}
\newcommand{\ba}{\begin{eqnarray}}
\newcommand{\ea}{\end{eqnarray}}
\newcommand{\baa}{\begin{eqnarray*}}
\newcommand{\btab}{\begin{tabular}}
\newcommand{\etab}{\end{tabular}}
\newcommand{\eaa}{\end{eqnarray*}}

\newcommand{\derleft}{\stackrel{\leftarrow}{D}\!}
\newcommand{\derright}{\stackrel{\rightarrow}{D}}

\def \labeltest #1 {\label{#1}}
\newcommand\re[1]{(\ref{#1})}

\newcommand\lr[1]{{\left({#1}\right)}}

\def\II{\hbox{{1}\kern-.25em\hbox{l}}}

\newcommand \vev [1] {\langle{#1}\rangle}

\def \CO {{\cal O}}

\begin{document}

\begin{titlepage}
\begin{flushright}
\begin{tabular}{l}
LPT--Orsay--00--04\\
SPbU--IP--00--02\\
TPR--00--01\\
hep-ph/0001130
\end{tabular}
\end{flushright}
\vskip1cm
\begin{center}
  {\large \bf
     Evolution of twist-three parton distributions in QCD beyond the
     large-$N_c$ limit
  \\}
\vspace{1cm}
{\sc V.M.~Braun}${}^{1}$,
{\sc G.P.~Korchemsky}${}^2$
          and {\sc A.N.~Manashov}${}^3$
\\[0.5cm]
\vspace*{0.1cm} ${}^1${\it
   Institut f\"ur Theoretische Physik, Universit\"at
   Regensburg, \\ D-93040 Regensburg, Germany
                       } \\[0.2cm]
\vspace*{0.1cm} ${}^3$ {\it
Laboratoire de Physique Th\'eorique%
\def\thefootnote{\fnsymbol{footnote}}%
\footnote{Unite Mixte de Recherche du CNRS (UMR 8627)},
Universit\'e de Paris XI, \\
91405 Orsay C\'edex, France
                       } \\[0.2cm]
\vspace*{0.1cm} ${}^4$ {\it
Department of Theoretical Physics,  Sankt-Petersburg State
University, \\ St.-Petersburg, Russia
                       } \\[1.0cm]

\vskip0.8cm
{\bf Abstract:\\[10pt]} \parbox[t]{\textwidth}{
We formulate a consistent $1/N_c^2$ expansion of the QCD evolution
equations for the twist-three quark distributions $g_2(x,Q^2)$,
$h_L(x,Q^2)$ and $e(x,Q^2)$ based on the interpretation of the
evolution as a three-particle
quantum-mechanical problem with hermitian Hamiltonian.
Each distribution amplitude can be decomposed in contributions of
partonic components with DGLAP-type scale dependence.
We calculate the $1/N_c^2$ corrections to the evolution of the
dominant component with the lowest anomalous dimension -- the only one
that survives in the large-$N_c$ limit -- and observe a good agreement
with the exact numerical results for $N_c=3$.
The $1/N_c^2$ admixture of operators with higher anomalous dimensions
is shown to be concentrated at a few lowest partonic components
and in general is rather weak.
}
\vskip1cm

\end{center}

\end{titlepage}

{\large\bf 1.~~}
Twist-three parton distributions in the nucleon are attracting
increasing interest as unique probes of novel quark-gluon correlations
in hadrons with clear experimental signature, giving rise to certain
asymmetries in experiments with polarized beams and targets.
Quantitative studies of such asymmetries are becoming possible
with the increasing precision of experimental data at SLAC and RHIC,
and can provide for an
important part of the future spin physics program on high-luminosity
accelerators like ELFE, upgraded CEBAF etc.
With this perspective, a detailed theoretical study of twist-three
parton distributions in QCD becomes mandatory.

Altogether, there exist three twist-3 distribution
functions \cite{JJ92} -- chiral-odd, $e(x,Q^2)$ and $h_L(x,Q^2)$,
and chiral-even, $g_2(x,Q^2)$ -- each being a function of parton momentum
fraction $x$ and the energy scale $Q^2$. By virtue of Lorentz invariance
the distributions $h_L$ and $g_2$ contain
contributions of twist-2 structure functions $h_1$ and $g_1$,
respectively:
\be
h_L(x) = 2x\int_{x}^1 \frac{dy}{y^2} h_1(y) + \widetilde h_L(x)
\,,\qquad
g_2(x) =-g_1(x) + \int_{x}^1 \frac{dy}{y} g_1(y) + \widetilde g_2(x)\,.
\ee
The QCD description of the remaining
genuine twist-3 part of these distributions
functions, $\widetilde h_L(x)$, $\widetilde g_2(x)$ and of $e(x)$,
is usually believed to be quite sophisticated. Their moments
are related through the QCD equations of motion to
the matrix elements of quark-antiquark-gluon
operators which have a nontrivial scale dependence and mix with
each other under the renormalization.%
\footnote{Additional mixture with three-gluon operators is present for
flavor-singlet contribution to $g_2(x,Q^2)$. In this letter
we consider only the flavor-nonsinglet part.}
As a consequence, the
QCD evolution equations for the 
functions $\widetilde h_L(x)$, $\widetilde g_2(x)$, $e(x)$
cannot be written in a closed form and require
additional nonperturbative input.
An important simplification occurs, however,
in the large-$N_c$ limit. It was shown \cite{ABH,BBKT} that to this
accuracy the twist-three distributions $\widetilde h_L$ and $e$
as well as the flavor-nonsinglet contribution to $\widetilde g_2$
satisfy simple DGLAP-type evolution equations (see e.g. \cite{GMR96})
\begin{eqnarray}
&&
  Q^2 \frac{d}{d Q^2} f(x,Q^2) = \frac{\alpha_s}{4\pi}
     \int_x^1\! \frac{dz}{z} P_f^{(0)}(x/z) f(z,Q^2)\,,
     \qquad f = \{\widetilde h_L\,,  \widetilde g_2^{_{\rm NS}}\,,e\}\,,
\nonumber\\
   && P_f^{(0)}(z) = 2 N_c\left\{
    \left[\frac{1}{1-z}\right]_+
      +\frac14\,\delta(1-z)
      +\sigma_f- \frac{1}{2}\right\},
\label{AP0}
\end{eqnarray}
where $\sigma_f = -1\,,0$ and $1$ for
$f =\widetilde h_L, \widetilde g_2$ and $e$,
respectively, and $1/(1-z)_+=1/(1-z)-\delta(1-z)\int_0^1
dz'/(1-z')$. This result implies that the inclusive
measurements of twist-three distributions are complete (to the
stated accuracy) in the sense that knowledge of the distribution
at one value of $Q^2$ is enough to predict the distribution at
arbitrary $Q^2$.
{}From phenomenological point of view it allows
to relate the measurements
of different experiments to each other and to
compare them to the model predictions (including lattice
calculations) that typically refer to a low scale.

Although the large-$N_c$ version of QCD presents a valid theoretical
limit, its relevance to the actual $N_c=3$ world and
its accuracy in predicting the scale dependence of twist-3
distributions is {\em a priory} not clear.
This work presents the first attempt to go beyond the large-$N_c$
approximation in a systematic way. In particular, we calculate the  $1/N_c^2$
corrections to the evolution kernels in (\ref{AP0}) and show that
mixing with quark-antiquark-gluon operators remains under control.

\vskip0.3cm

{\large\bf 2.~~}
From the OPE analysis \cite{SV82} one finds that
the scale dependence of the moments of the
twist-3 distributions $\int_{-1}^1 dx x^{N+2}f(x)$
is governed by renormalization of the set of local 
composite quark-antiquark-gluon operators ($k=0$, $...$, $N$)
\begin{eqnarray}
[S^{\pm}_{\mu}]^k_N &=& \bar q (\derleft\cdot n)^k \not\! n\,
    n^\nu [\widetilde G_{\mu\nu}\pm i  G_{\mu\nu}\gamma_5]
\,(\derright\cdot n)^{N-k}q\,,
\nonumber\\{}
[T_\Gamma]_N^k &=& \bar q (\derleft\cdot n)^k n_\mu
\sigma^{\mu\rho}\,
\Gamma\, n^\nu G_{\nu\rho}\,(\derright\cdot n)^{N-k}q\,,\qquad
\Gamma=\{\II,\,i\gamma_5\}
\label{local}
\end{eqnarray}
for chiral-even ($S^{\pm}_{\mu}$) and chiral-odd distributions
($T_{\rm I}$ and $T_{i\gamma_5}$ for $e(x)$ and $\widetilde h_L(x)$,
respectively), see e.g. \cite{BKL84}.
Here, $n_\mu$ is a light-like vector and
$\widetilde G_{\mu\nu}=\epsilon_{\mu\nu\rho\lambda}
G^{\rho\lambda}/2$ stands for a dual gluon field strength.
To leading order, renormalization of $T_{\rm I}$ and $T_{i\gamma_5}$
is the same and we, therefore, drop the subscript in what follows.
Similarly, it is enough to consider the  operator $S^+$.

The operators $[T]_N^k$ (and $[S^+]_N^k$)
with different $k=0,...,N$ and the same number of covariant derivatives
$N\ge 0$ mix with each other under renormalization.
The mixing matrices have been calculated to the leading order
(e.g. \cite{BKL84}) and can be diagonalized numerically
for any given $N$. The eigenvectors, then,
define the multiplicatively renormalizable operators and the eigenvalues give
 their anomalous dimensions.
A disadvantage of this (traditional) approach is that the mixing
matrix does not have any obvious structure in this basis
and is not symmetric. As the result, the structure of the spectrum
remains obscure and the
eigenvectors are not mutually orthogonal with any simple weight function.
Relative importance of various contributions is also far from being
clear.%
\footnote{E.g. the hierarchy of entries in the mixing matrix
$\CO(N)$, $\CO(1)$, $\CO(1/N)$, etc. is not preserved in the eigenvalues.}

A systematic $1/N_c^2$ expansion of the evolution equations is made
possible by going over to the Hamiltonian formulation developed in
in \cite{BFLK,BDM98,BDKM,Belitsky99,DKM99}. The renormalization
group evolution is driven to leading logarithmic accuracy by tree-level
counterterms and has the conformal symmetry of the QCD Lagrangian.
The evolution kernel can, therefore, be written in an abstract
operator form in terms of Casimir operators of the collinear subgroup
$SL(2,R)$ of the conformal group. In this way, diagonalization of the
mixing matrix of the twist-3 operators in (\ref{local}) can be
reformulated as a three-particle quantum mechanical problem
\begin{equation}
{\cal H}\,\Psi_{N,q}(x_1,x_2,x_3)={\cal E}_{N,q}\Psi_{N,q}(x_1,x_2,x_3),
\label{solution}
\end{equation}
defined by the Hamiltonian \cite{BFLK,BDM98,Belitsky99,DKM99}
\begin{equation}
   {\cal H}_A={N_c}{\cal H}^{(0)}_A-\frac{2}{N_c}{\cal
   H}^{(1)}_A\,,\qquad A=\{T\,, S^+\}\,,
\label{H}
\end{equation}
with
\begin{eqnarray}
\label{exact}
   {\cal H}^{(0)}_T &=& V_{qg}^{(0)}(J_{12}) +
   V_{qg}^{(0)}(J_{23})\,,\qquad
   {\cal H}^{(1)}_T = V_{qg}^{(1)}(J_{12}) + V_{qg}^{(1)}(J_{23})
 + V_{qq}^{(1)}(J_{13})\,,
\nonumber\\
 {\cal H}^{(0)}_{S^+} &=& V_{qg}^{(0)}(J_{12}) +
 U_{qg}^{(0)}(J_{23})\,,\qquad
 {\cal H}^{(1)}_{S^+} = V_{qg}^{(1)}(J_{12}) + U_{qg}^{(1)}(J_{23})
 + U_{qq}^{(1)}(J_{13})\,.
\end{eqnarray}
Here, the notation was introduced for two-particle quark-quark and
quark-gluon kernels
\begin{eqnarray}
V_{qg}^{(0)}(J)&=&\psi(J+3/2)+\psi(J-3/2)-2\psi(1)-3/4\,,
\label{nonplanar}
\\
U_{qg}^{(0)}(J)&=&\psi(J+1/2)+\psi(J-1/2)-2\psi(1)-3/4\,,
\nonumber\\
V_{qg}^{(1)}(J)&=&\frac{(-1)^{J-5/2}}{(J-3/2)(J-1/2)(J+1/2)}\,,
\quad
U_{qg}^{(1)}(J)=-\frac{(-1)^{J-5/2}}{2(J-1/2)}\,,
\nonumber\\
V_{qq}^{(1)}(J)&=&\psi(J)-\psi(1)-3/4\,,
\quad
U_{qq}^{(1)}(J)=\frac12\left[\psi(J-1)+\psi(J+1)\right]-\psi(1)-3/4\,.
\nonumber
\end{eqnarray}
Here and below $\psi(x)=d\ln\Gamma(x)/d x$ stands for the Euler
$\Psi$-function; the subscripts `1,2,3' refer to antiquark,
gluon and quark fields, respectively.
The operators $J_{ik}$  are defined as follows
\begin{equation}
J_{ik}\,(J_{ik}-1)=L_{ik}^2
=(\vec{ L}_i+\vec{L}_k)^2\,,
\label{Jik}
\end{equation}
where $\vec{L}_i$ are the generators of the $SL(2,R)$ group and $L_{ik}^2$
are the corresponding two-particle Casimir operators.
The generators $\vec{L}_i$ can be realized as differential operators
acting on coordinates $x_i$ of the wave function $\Psi(x_1,x_2,x_3)$
\begin{equation}
L_{-,\,i}=x_i\partial^2_{x_i}+2j_i\partial_{x_i}\,,~~~~
L_{+,\,i}= -x_i\,,~~~~
L_{0,\,i}=x_i\partial_{x_i}+j_i\,,
\label{S_i}
\end{equation}
where the conformal spin $j$
is equal to $j_1=j_3=1$ for the (anti)quark and $j_2=3/2$
for the gluon field, respectively. The eigenvalues of the operator $J_{ik}$
define the possible values of the conformal spin in the
two-parton channel and are given by $j_i+j_k+n$ with $n$ being nonnegative integer.

The Hamiltonians defined above are manifestly $SL(2,R)$ invariant:
\begin{equation}
[{\cal H}, L_\alpha]=[{\cal H}, L^2]=[L^2,L_\alpha]=0,
\end{equation}
where $L_\alpha$ ($\alpha=0,+,-$) are the total three-particle
$SL(2)$ generators
\begin{equation}
L_\alpha=L_{\alpha,1}+L_{\alpha,2}+L_{\alpha,3}\,,~~~~
L^2=L_0(L_0-1) + L_+L_-\,.
\end{equation}
One can, therefore, diagonalize the three operators $L^2,L_0$ and
$\cal{H}$ simultaneously. The  conditions
\begin{equation}
L_0\,\Psi_{N,q}(x_1,x_2,x_3)=\left(N+7/2\right)
\Psi_{N,q}(x_1,x_2,x_3)\,,
\qquad
L_{-}\,\Psi_{N,q}(x_1,x_2,x_3)=0
\label{constraint}
\end{equation}
define $\Psi_{N,q}(x_1,x_2,x_3)$ to be a homogeneous polynomial of
degree $N$ that does not contain factors of $(x_1+x_2+x_3)$ and therefore
does not vanish as $\sum_i x_i=0$.

Solving the Schr\"odinger equation \re{solution} one constructs the basis of
multiplicatively renormalizable local operators, ${\cal O}_{N,q}$.
Omitting the Lorentz structure, the correspondence is,
schematically~\cite{BDKM}
\begin{equation}
{\cal O}_{N,q} = \Psi_{N,q}(D_{\bar q},D_g,D_q)\, \bar q\, G\, q
\label{O-Nq}
\end{equation}
with covariant derivatives $D_{\bar q},D_g,D_q$ acting on the antiquark, gluon
and quark, respectively.
The corresponding eigenvalues provide
the anomalous dimensions ${\cal E}_{N,q}$:
\begin{equation}
   {\cal O}_{N,q}(Q^2) =
\left(\frac{\alpha_s(Q^2)}
{\alpha_s(\mu^2)}\right)^{{\cal E}_{N,q}/b}\!\!
{\cal O}_{N,q}(\mu^2)\,,
\label{renorm}
\end{equation}
where $b= 11 N_c/3-2 n_f/3$. Here, parameter $q$ enumerates
different eigenstates of the mixing matrix.

Note that the conformal operators defined in (\ref{O-Nq})
diagonalize the full mixing matrix, including mixing with the operators
containing total derivatives.  Since at the end only forward matrix elements
of these operators enter the parton distributions, taking into account
this additional mixing may be seen as an unnecessary complication.
The advantage for such formulation is, however, that
the Hamiltonian becomes hermitian with respect to the scalar product
\be
\vev{\Psi_1 |\Psi_2} = \int_0^1 {\cal D}x \, x_1 x_2^2 x_3
\Psi_1^*(x_i)\Psi_2(x_i)
\label{prod}
\ee
with ${\cal D}x=dx_1dx_2dx_3\delta(x_1+x_2+x_3-1)$. As well known in
quantum mechanics, hermiticity of the Hamiltonian  implies that all its
eigenvalues
(anomalous dimensions) ${\cal E}_{N,q}$ are real\footnote{
Although the conformal symmetry is lost beyond one-loop, it is
easy to prove that anomalous dimensions of twist-3 operators remain real
to all orders in perturbation theory. Indeed, entries in the
mixing matrix are real and complex eigenvalues may only appear in
(complex conjugate) pairs. On the other hand,
the number of eigenvalues is fixed and in leading order
they are non-degenerate (which one can establish
by inspection). These two conditions are contradictory since
a pair of complex conjugate eigenvalues (i.e. with the same real parts)
cannot be obtained from a discrete non-degenerate
spectrum at $\alpha_s\to0$ (i.e. with all real parts different)
by a perturbative renormalization group flow.},
and the eigenfunctions are mutually orthogonal:
\be
  \vev{\Psi_{N,q} |\Psi_{N',q'}} =
  \delta_{NN'}\delta_{qq'}|\!|\Psi_{N,q}|\!|^2,
  \qquad
  |\!|\Psi_{N,q}|\!|^2 =
  \int_0^1 {\cal D}x \, x_1 x_2^2 x_3
  |\Psi_{N,q}(x_i)|^2\,.
\label{sc-prod}
\ee
Completeness and orthogonality of the eigenstates corresponding to
multiplicatively renormalizable operators will be of crucial
importance for this work. In particular, using these properties
one can expand the moments of the twist-3 parton distributions
over the contributions of multiplicatively renormalizable operators as
\be
\int_{-1}^1 \!dx\,x^{N+2}\,f(x,Q^2) =
\sum_{q}\frac{\langle \Psi_{N,q}|\Phi_{N}\rangle}
{|\!|\Psi_{N,q}|\!|^2}
\langle\!\langle \CO_{N,q}(Q^2)\rangle\!\rangle\,.
\label{decomp}
\ee
Here, $\langle\!\langle \CO_{N,q}\rangle\!\rangle$
are reduced (scalar and dimensionless) forward
matrix elements of the corresponding local operators over the nucleon
state and $\Phi_{N}(x_i)$  are the
coefficients in the re-expansion of twist-3
quark-antiquark operators that define the parton distributions, in
terms of quark-antiquark-gluon operators.
They are given by QCD equations of motion and for
$x_1+x_2+x_3=0$ take the simple form~\cite{DKM99}
\begin{equation}
   \Phi^{T,{+}}_N =
       \frac{x_1^{N+1}-(-x_3)^{N+1}}{x_1+x_3}\,,~~
   \Phi^{T,{-}}_N =
      (\partial_{x_1}+\partial_{x_3}) \frac{\Phi^{T,{+}}_{N+1}}{N+2}\,,~~
   \Phi^{S^+}_N = \partial_{x_1}\frac{\Phi^{T,{+}}_{N+1}}{N+2}\,
\label{WF1}
\end{equation}
for $f=e\,, \widetilde h_L$ and $\widetilde g_2$, respectively.
Note, however, that in order to form the scalar product in (\ref{decomp})
one needs to know $\Phi_N(x_i)$ for $x_1+x_2+x_3=1$ that
corresponds to taking into account contributions to the QCD equations
of motion of the operators containing total derivatives.
 The corresponding
expressions are given below in (\ref{expandcoef}).
Each term in \re{decomp} has an autonomous $Q^2-$evolution,
Eq.~\re{renorm}, and brings a new nonperturbative parameter
$\langle\!\langle \CO_{N,q}(\mu^2)\rangle\!\rangle$.

\vskip0.3cm

{\large\bf 3.~~}
The large-$N_c$ Hamiltonians ${\cal H}^{(0)}$ in (\ref{exact}) possess an
additional `hidden' symmetry, related to the existence of a nontrivial
conserved charge \cite{BDM98}
\begin{eqnarray}
  &&[{\cal H}^{(0)}_{T},Q_T] =0\,, ~~~~
  Q_T = \{L_{12}^2,L_{23}^2\} -\frac92 L_{12}^2 - \frac92 L_{23}^2\,,
\nonumber\\{}
  &&[{{\cal H}}^{(0)}_{S^+},Q_{S^+}] =0\,, ~~~~
  Q_{S^+} = \{L_{12}^2,L_{23}^2\} - \frac12 L_{12}^2 - \frac92 L_{23}^2\,,
\label{Q}
\end{eqnarray}
where $\{~,~\}$ stands for the anticommutator. As the result,
eigenstates of  ${\cal H}^{(0)}$ can be labeled by quantized eigenvalues
$q_\ell$, \, $(\ell = 0,1,\ldots,N)$,
of the operator $Q$:
\begin{equation}
   Q\, \Psi_{N,q}^{(0)}(x_1,x_2,x_3) = q\, \Psi_{N,q}^{(0)}(x_1,x_2,x_3)\,.
\label{Q-eq}
\end{equation}
Since the number of degrees of freedom (=3) equals in this case to
the number of conserved charges ($L^2, L_0, Q$), the quantum mechanical system
described by the Hamiltonian  ${\cal H}^{(0)}$ is completely integrable.
This allows to use advanced mathematical methods for the analysis of
the spectrum and, in particular, develop the WKB expansion of
the eigenvalues at large $N$ \cite{BDM98,Belitsky99,DKM99}.

The remarkable property of the large-$N_c$ evolution kernels \cite{ABH,BBKT}
that is responsible for the simplicity of the evolution in (\ref{AP0})
is that the eigenfunctions corresponding to the states with lowest
energy coincide with the coefficient functions $\Phi_N$
entering the expansion in Eq.~(\ref{decomp}) \cite{Belitsky99,DKM99}
\be
\Phi_{N}(x_i)=\Psi_{N,q_0}^{(0)}(x_i)\,,\quad
q_0^{S}=(N+3)^2+3/8\,,\quad
q_0^{T,\pm}=(N+3\mp 2)^2-53/8\,.
\label{Psi0}
\ee
To be precise, $\Phi^{S^+}_{N}(x_i)$ coincides with
the ground state of ${\cal H}^{(0)}_{S^+}$ and
$\Phi^{T,\pm}_{N}(x_i)$ with the
two lowest eigenstates of ${\cal H}^{(0)}_T$ with opposite parity
with respect to permutations of quarks, $x_1\rightleftarrows x_3$.
 The corresponding eigenvalues
determine the lowest anomalous dimensions
for each $N$, ${\rm min}_q\, {\cal E}_{N,q}={\cal E}_{N,q_0}=N_c E_N
+{\cal O}(1/N_c)$ and they  are equal to
\begin{eqnarray}
&&E_N^{T,{\pm}} = 2\psi(N+3)+\frac{1\mp 2}{N+3} -\frac12+2\gamma_{_{\rm E}}\,,
\nonumber\\
&&E_N^{S} = 2\psi(N+3)+\frac1{N+3} -\frac12+2\gamma_{_{\rm E}}\,.
\label{E-ex}
\end{eqnarray}
We will often omit the subscript ``$q=q_0$'' for these special states.
Note useful relations:
\begin{equation}
2E_N^{S}= E_N^{T,{+}}+E_N^{T,{-}}\,,
~~
2\Phi_N^{S^\pm}(x_i)=\Phi_N^{T,{+}}(x_i)\pm\Phi_N^{T,{-}}(x_i)\,,
~~
\vev{\Phi_N^{T,{-}} |\Phi_N^{T,{+}}} = 0\,.
\label{cor0}
\end{equation}
Since the wave functions of eigenstates with different energies are
mutually orthogonal, Eq.~\re{Psi0} implies that the single term
$q=q_0$ survives in the sum \re{decomp} in the leading large $N_c$ limit.
As a consequence, the moments of twist-3 parton distributions are entirely given
by the reduced matrix elements of the
corresponding local operators $\CO_{N,q_0}$, Eq.~(\ref{O-Nq})
\cite{ABH,BBKT}:
\begin{eqnarray}
  \int_{-1}^1 \!dx\,x^{N+2}\,e(x,Q^2) &=&
\langle\!\langle\CO_N^{T,{+}}(Q^2)\rangle\!\rangle\,,
\nonumber\\
(N+4) \int_{-1}^1 \!dx\,x^{N+2}\,\widetilde h_L(x,Q^2) &=&
\langle\!\langle\CO_N^{T,{-}}(Q^2)\rangle\!\rangle\,,
\\
\frac{N+3}{N+2}
\int_{-1}^1 \!dx\,x^{N+2}\,\widetilde g_2(x,Q^2) &=&
   \langle\!\langle\CO_N^{S^+}(Q^2)\rangle\!\rangle
    +\langle\!\langle\CO_N^{S^-}(Q^2)\rangle\!\rangle
    \,. \nonumber
\end{eqnarray}
Combined with the scale dependence of matrix elements, Eq.~(\ref{renorm}),
these relations are equivalent to the DGLAP evolution equations in (\ref{AP0})
with $N_cE_N^{f}=-\int_0^1 dz\, z^{N+2} P_f^{(0)}(z)$.

Treating the $\sim 1/N_c$ contribution  to the Hamiltonian in (\ref{H})
as a  perturbation, we expand
\begin{eqnarray}
  {\cal E}_{N,q} &=& N_c E_{N,q} + N_c^{-1} \delta E_{N,q}+\ldots\,,
\label{momN}\\
  \Psi_{N,q} &=& \Psi^{(0)}_{N,q} +  N_c^{-2}\delta \Psi_{N,q}+\ldots\,,
\nonumber
\end{eqnarray}
with the usual quantum-mechanical expressions
\begin{eqnarray}
  \delta E_{N,q} &=& -2 |\!|\Psi^{(0)}_{N,q}|\!|^{-2}\,
\vev{\Psi^{(0)}_{N,q}|{\cal H}^{(1)}|\Psi^{(0)}_{N,q}}\,\,,
\label{deltaE}\\
   \delta \Psi_{N,q}(x_i) &=& -2\sum_{q'\not= q}
   \frac{\vev{\Psi^{(0)}_{N,q}|{\cal H}^{(1)}|\Psi^{(0)}_{N,q'}}}
        {|\!|\Psi^{(0)}_{N,q'}|\!|^2}
        \frac{\Psi^{(0)}_{N,q'}(x_i)}{{E_{N,q}-E_{N,q'}}}\,.
\label{deltaPsi}
\end{eqnarray}
To this accuracy, moments of the twist-3 distributions
are no longer renormalized multiplicatively and have to be expanded in
contributions of multiplicatively renormalizable operators, Eq.~\re{decomp}:
\begin{equation}
  c_f(N)\int_{-1}^1 \!dx\,x^{N+2}\,f(x,Q^2) =
    \langle\!\langle \CO_{N,q_0}(Q^2)\rangle\!\rangle
-\frac{2}{N_c^2} \sum_{q\not= q_0}
    \frac{\langle \Psi^{(0)}_{N,q}|{\cal H}^{(1)}|\Phi_{N}\rangle}
        {|\!|\Psi^{(0)}_{N,q}|\!|^2}
\cdot
        \frac{\langle\!\langle \CO_{N,q}(Q^2)\rangle\!\rangle}
             {{E_{N,q}-E_{N,q_0}}}\,,
\label{mixing}
\end{equation}
where $c_e =1$, $c_{h_L}= N+4$, $c_{g_2} = (N+3)/(N+2)$,
$\Phi_{N}\equiv \Psi_{N,q_0}^{(0)}$ is one of the
functions $\Phi_N^{S^+}$, $\Phi_N^{T,{+}}$, $\Phi_N^{T,{-}}$.
$\CO_{N,q_0}(Q^2)$ is the quark-gluon-antiquark operator
with the lowest anomalous dimension corresponding to the wave function
$\Psi_{N,q_0} =\Psi_{N,q_0}^{(0)}+N_c^{-2}\delta \Psi_{N,q_0}$ and
normalized at the scale $Q^2$.
Thus, the calculation of $1/N_c^2$ corrections to the moments of the
structure functions consists in two separate tasks.
First, one has to calculate the $\CO(1/N_c^2)$
corrections to the anomalous dimension of $\CO_{N,q_0}$, or
equivalently to the energies (\ref{E-ex})
of the lowest eigenstates and, second, calculate the mixing coefficients
${\langle \Psi^{(0)}_{N,q}|{\cal H}^{(1)}|\Psi_{N,q_0}^{(0)}\rangle}$
with higher levels; evolution
of the latter can be taken into account in the leading large-$N_c$
approximation of Refs.~\cite{Belitsky99,DKM99}.
We address both questions  in what follows.
\vskip0.3cm

{\large\bf 4.~~}
The $\CO(1/N_c^2)$ correction $\delta E_N$ to the energy
of the lowest eigenstates is given by Eq.~(\ref{deltaE}) for $q=q_0$. The calculation
is most easily done by going over to the so-called conformal basis
\cite{BDM98,BDKM,Belitsky99,DKM99}. The idea is to define a
basis of (orthogonal with respect to \re{prod}) polynomials
that satisfy the conformal constraints in (\ref{constraint}) and, in addition,
diagonalize the two-particle Casimir operator (\ref{Jik}) in one of the channels, e.g.
\begin{equation}
   L_{12}^2\, Y^{(12)3}_{N,n}(x_1,x_2,x_3) = (n+5/2)(n+3/2)\,Y^{(12)3}_{N,n}(x_1,x_2,x_3)
\label{Y}
\end{equation}
and, similarly, $Y^{1(23)}_{N,n}$ and $Y^{(13)2}_{N,n}$.
Here, the superscripts indicate the order in which the conformal spins
of the three partons
sum up to the total conformal spin $N+7/2$. The three
different sets of $Y$-functions are related to each other through the
Racah $6j$-symbols of the $SL(2)$ group. Permutation symmetry
between the quarks implies that
$Y^{(23)1}_{N,n}(x_1,x_2,x_3)=(-1)^n\,Y^{(12)3}_{N,n}(x_3,x_2,x_1)$.
The explicit expressions
for the $Y-$functions are given in terms of the Jacobi polynomials
\cite{BDKM,Belitsky99}.

The expansion coefficients of the lowest eigenstates  (\ref{WF1})
in each of the three conformal basis are easily obtained
\cite{DKM99,2000}
\ba
\Phi_N^{T,\pm}&=&\sum_{n=0}^N (-1)^n(N+n+5)
\left[\frac{N+3}{n+1}\pm(n+3)\right]Y^{(12)3}_{N,n}
\\
&=&\sum_{n=0}^N \left[(-1)^{N-n}\pm 1\right](2n+3)(n+2)\frac{2(N+n)+9\pm 1}
{2(N-n)+3\mp 1}Y^{(31)2}_{N,n}
\nonumber
\label{expandcoef}
\ea
and $\Phi_N^{S^+}=(\Phi_N^{T,+}+\Phi_N^{T,-})/2$.
The normalization is
\ba
|\!|Y^{(12)3}_{N,n}|\!|^2&=&\frac{(n+1)(N-n+1)}{(n+2)(n+3)(N+n+5)}\,,
\nonumber\\
|\!|Y^{(31)2}_{N,n}|\!|^2&=&\frac{2(n+1)(N-n+1)(N-n+2)}
{(n+2)(2n+3)(N+n+4)(N+n+5)}\,.
\ea
Notice that the two-particle kernels defined in \re{nonplanar}
become diagonal in the corresponding basis. Expanding each contribution
in a suitable basis,
one obtains the matrix elements,
$\vev{\Psi_N^{T,\pm}|{\cal H}^{(1)}_T|\Psi_N^{T,\pm}}$ and
$\vev{\Psi_N^{S^+}|{\cal H}^{(1)}_{S^+}|\Psi_N^{S^+}}$,
as finite sums over two-particle spins $n$.
Full expressions are rather cumbersome and will
be presented elsewhere \cite{2000}.

Expanding the resulting expressions for $\delta E_N$ in powers of $1/(N+3)$
we obtain
\ba
\delta E^{S}_N&=&
    - 2\left(\ln(N+3)+\gamma_{_{\rm E}}+\frac34 -
    \frac{\pi^2}{6}\right) + \CO\left(\frac{\ln^2 (N+3)}{(N+3)^2}\right)\,,
\label{Energy1}\\
\delta E^{T,{\pm}}_N
& =&
\delta E^{S}_N
\pm~\frac4{N+3}\left[ \left(3-\frac{\pi^2}{3}\right)
\lr{\ln(N+3)+ \gamma_{_{\rm E}}}
-\frac{5}{2}+\frac{\pi^2}{3}\right]+\CO\left(\frac{\ln^2 (N+3)}{(N+3)^2}\right)\,.
\nonumber
\ea
With this correction, Eq.~(\ref{momN})
gives an excellent description of the lowest anomalous
dimension in the spectrum of twist-3 operators for
{\it all\/} integer $N\ge 0$, see Table~1.
Note that to this accuracy  ${\cal E}^{T,{+}}_N+{\cal E}^{T,{-}}_N
= 2{\cal E}^{S}_N$, cf. Eq.~(\ref{cor0}).
\def\thefootnote{}\footnote{\hspace*{-3mm}${}^\fnsymbol{footnote}$%
         The difference of these values with the exact result
         is entirely due to the truncation of the $1/(N+3)-$expansion
         in (\ref{Energy1}). Since only one independent operator exists
         in these three cases, there is no mixing and the $\CO(1/N_c)$ approximation to
         the energies ${\cal E}^{S}_{N=0}$, ${\cal E}^{T,{+}}_{N=0}$ and
         ${\cal E}^{T,{-}}_{N=1}$ is in fact exact.}%
\def\thefootnote{\fnsymbol{footnote}}%

The following comments are in order.

\begin{table}[t]
\begin{center}
\begin{tabular}{|c||c|c|c|c|c|c|c|}
\hline
$N$ & $0$ &  $1$  & $2$ & $10$ & $20$ &$50$ &$100$ \\
\hline
\hspace*{1.2cm}  {\rm exact}
&$6.1111$  & $8.1111$  &$9.5902$  &$15.3901$ &$18.6260$
         &$23.2234 $& $26.8234 $ \\
${\cal E}^{T,{+}}_N$:\hspace*{0.3cm} Eq.\re{Energy1}
&
\hspace*{1.0mm}$6.1146^\fnsymbol{footnote}$\hspace*{-1.0mm}
&  $8.1112$ &$9.5805$
&$15.3779$ &$18.6193$
         &$23.2233 $ & $26.8260 $\\
\hspace*{1.2cm} large-$N_c$ &$6.5000$  &  $8.7500$ &$10.4000$
&$16.8885$ &$20.5144$
         &$25.6717 $ & $29.7134 $\\
\hline
\hspace*{1.2cm} {\rm exact} &     --   &$11.5556$
&$12.2111$  &$16.3820$ &$19.1948$
         &$23.4791 $ & $26.9596 $ \\
${\cal E}^{T,{-}}_N$:\hspace*{0.3cm}  Eq.\re{Energy1} 
&     --   &
\hspace*{1.0mm}$10.9640^\fnsymbol{footnote}$\hspace*{-1.0mm}
&$11.8972$
&$16.3258$ &$19.1742$
         &$23.4763 $ &$ 26.9612 $ \\
\hspace*{1.2cm} large-$N_c$ &     --   &$11.7500$   &$12.8000$
&$17.8116$ &$21.0362$
         &$25.8981 $ &$ 29.8299 $ \\
\hline
\hspace*{1.2cm} {\rm exact}
&$8.5556 $ & $ 9.7550$ &$10.8914 $ &$15.8758 $ &$18.9033
         $ &$23.3480  $& $26.8932 $ \\
${\cal E}^{S}_N$: \hspace*{0.4cm}  Eq.\re{Energy1} 
&
\hspace*{1.0mm}$7.9794^\fnsymbol{footnote}$\hspace*{-1.0mm}
&$ 9.5376$&$ 10.7389 $&$
15.8519 $&$ 18.8968 $&$ 23.3498 $&$ 26.8936$ \\
\hspace*{1.2cm} large-$N_c$ &$ 8.5000 $& $10.2500 $ &$11.6000 $&$
17.3500 $&$ 20.7753$ &$ 25.7849 $ & $29.7716 $\\
\hline
\end{tabular}
\end{center}
\caption{The lowest anomalous dimensions in the spectrum of twist-3
operators for different $N$ calculated by taking into account the $\CO(1/N_c)$
correction, Eqs.~(\ref{Energy1}) and \re{momN}, in comparison with the
corresponding exact numerical results and leading large-$N_c$ expressions \re{E-ex}.}
\label{Table1}
\end{table}

First, we note that $\delta E_N$ has the same
large-$N$ behavior, $\delta E_N\sim\ln N$, as the leading large-$N_c$
result in \re{E-ex}.
The coefficient
in front of the $\ln N$ term at large $N$ is redefined, therefore, from
$2N_c$ to $4C_F=2(N_c^2-1)/N_c$, in agreement with \cite{ABH,BBKT}.
It is possible to show \cite{K89} that this coefficient is exact and
higher order $1/N_c^2$ corrections do not grow with $N$,
${\cal E}_{N,q_0} = 2 C_F\ln N + \CO(N^0)$.

Second, the constant $\CO(N^0)$ term in the large $N$
expansion of the anomalous dimensions \re{Energy1}
does not agree with \cite{ABH,BBKT,GMR96}.
The difference is due to the contribution of the operators $V_{qg}^{(1)}$ and
$U_{qg}^{(1)}$ in (\ref{exact}) that was overlooked in the previous works. The
result for the $\CO(1/N)$ correction is new.

{}Finally, from the expressions in (\ref{Energy1}) it is easy to read out the
corresponding modification of the DGLAP splitting functions
(\ref{AP0}), $P_{f}(z)= P^{(0)}_{f}(z)+P^{(1)}_{f}(z)$. For example
\be
P^{(1)}_{g_2}(z)
\stackrel{z\to1}{=}
\frac{2}{N_c} \left\{\left[ \frac{-1}{1-z}\right]_{+}\!\!+
\left(\frac34-\frac{\pi^2}{6}\right)\delta(1-z)+\frac12
+\CO(1-z) \right\}.
\label{AP1}
\ee
This result is exact to the ${\cal O}(1/N_c^4)$ accuracy and neglecting
all terms that vanish at $z\to1$. Expressions
for $P^{(1)}_{e}(z)$ and $P^{(1)}_{h_L}(z)$ have similar structure.

The opposite limit of the small$-z$ behavior of the
DGLAP splitting functions $P_{f}(z)$ is of special interest.
It is well known that this
behavior is governed by singularities of the anomalous dimensions
${\cal E}_{N}$ in the complex $N$-plane. In particular,
 the leading large-$N_c$ anomalous dimensions (\ref{E-ex}) have
simple poles at negative integer $N\le -3$. The  pole at $N=-3$
corresponds to $P^{(0)}(z)\stackrel{z\to 0}{\sim} z^0$ in (\ref{AP0}).
With this connection in mind, we have studied the analytic structure
of $1/N_c^2$ corrections to the anomalous dimensions, $\delta E_N$,
continued analytically to the complex $N$ plane. The calculation
is straightforward, albeit tedious
\cite{2000}. We find that the analytic continuation has to be performed
separately for even and odd $N$, as familiar from studies of the evolution
of twist-2 parton distribution beyond the leading order \cite{CFP80}.
Similarly to the latter case, we define%
\footnote{This corresponds to the
decomposition over partial waves with  definite signature.}
\begin{equation}
\int_{-1}^1\!dx\, x^N \,f(x) =
\frac{1+(-1)^N}{2} \int_{0}^1\!dx\, x^N \,f_{\rm even}(x)+
\frac{1-(-1)^N}{2} \int_{0}^1\!dx\, x^N \,f_{\rm odd}(x)\,
\end{equation}
and consider the evolution of even- and odd-signature component of
$f= e\,,\widetilde h_{L}\,,\widetilde g_{2}$
separately. When the restriction to even (odd) $N$ is imposed,
the normalized matrix elements
$\vev{\Phi_{N}|{\cal H}^{(1)}|\Phi_{N}}/|\!|\Phi_{N}|\!|^{2}\,$
develop singularities at, generically,  $N=0,-1,-2$. With one exception, all
singularities to the right of $N=-3$ are simple poles and appear due to
vanishing of the norm of the leading large-$N_c$ wave functions
$|\!|\Phi_{N}|\!|^2$. They give rise to small-$z$ behavior of the
$1/N_c^2$ correction to the DGLAP splitting functions of the form
\ba
P^{(1)}_{g_2^{\rm even}}(z)&\stackrel{z\to0}{=}&
\frac1{N_c}\left\{
     -5\,\frac{17-2\pi^2}{51-4\pi^2}\frac{1}{z^2} +6 +{\cal O}(z)
\right\}
     \,,
\nonumber\\
P^{(1)}_{g_2^{\rm odd}}(z)&\stackrel{z\to0}{=}&
\frac1{N_c}\left\{
\frac{6}{6-\pi^2}\frac{\ln z}{z}
+\frac{\pi^4-18\pi^2+54+72\,\zeta(3)}{2(6-\pi^2)^2}\,\frac1z+\CO(z)
\right\}
\,.
\label{small-z}
\ea
This asymptotics is more singular than that of $P^{(0)}_{g_2}(z)$ and
is in apparent contradiction with the Regge theory expectations.
The origin of these ``spurious'' singularities and their physical significance
is not clear and deserves a special study that goes beyond the tasks
of this letter. We would like to stress that appearance of
singularities of the anomalous dimensions to the right from $N=-3$ can well be
artifact of the $1/N_c$ expansion, since close to the
singularities the $1/N_c^2$ correction dominates over
leading order term in \re{momN} and the $1/N_c^2$ expansion
formally breaks down.

Another delicate point is that analytic continuation in $N$ has to be done
in \re{mixing} simultaneously with the analytic continuation of the sum
over anomalous dimensions ${\cal E}_{N,q_\ell}$ belonging to
different ``trajectories'' parameterized by integer $\ell$ \cite{DKM99}.
To the best of our knowledge, the problem of analytic continuation
from a set of discrete points ${\cal E}_{N,q}$ on a $(N,q)-$plane has
never been addressed in mathematical literature.

To illustrate that the problem is nontrivial, consider the trace of the
(full) Hamiltonian \re{H} over the subspace spanned by wave functions with
given $N\ge 0$
\begin{equation}
 {\rm Tr}_N\, {\cal H} = \sum_{\ell=0}^{N} {\cal E}_{N,q_\ell}\,.
\end{equation}
It is equal to the sum of all $N+1$ anomalous
dimensions and is easily calculated {\em exactly} as a sum of
diagonal matrix elements in any suitable basis. We obtain,
in particular
\begin{eqnarray}
  {\rm Tr}_N\, {\cal H}^{(0)}_{S^+} &=&
  2(2N+5)\Big[\psi(N+3)+\gamma_{_{\rm E}}\Big]
 -\frac{11}{2}N -13 +\frac{1}{N+2}+\frac{1}{N+3}\,,
\nonumber\\
  {\rm Tr}_N\, {\cal H}^{(1)}_{S^+} &=&
(N+2)\Big[\psi(N+3)+\gamma_{_{\rm E}}\Big]
-\frac{7}{4}N -5 +\frac{1}{2(N+2)}
+\frac52\ln 2
\\
&&{}+(-1)^{N}\left\{
\frac54\left[\psi\left(\frac{N}{2}+\frac32\right)-\psi\left(\frac{N}{2}
  +2\right)\right]+\frac{1}{2(N+2)}+\frac{1}{2(N+3)}
\right\}.
\nonumber
\label{trace}
\end{eqnarray}
Notice now  that the trace of already the leading large-$N_c$ Hamiltonian
${\cal H}^{(0)}$ is singular at $N=-2$.
This implies that either one (or several)
trajectories of the anomalous dimensions, ${\cal E}_{N,q_\ell}$,
becomes singular at this point, or the (analytically continued) sum over
the trajectories in \re{mixing} diverges. In both cases, analytic continuation
of the lowest trajectory (\ref{E-ex}) beyond $N=-2$ is  questionable and,
therefore, asymptotic expressions in (\ref{small-z}) (or, equivalently,
the DGLAP-evolution for $z < 1/N_c^2$) have to be taken with caution.
\vskip0.3cm

{\large\bf 5.~~}
According to \re{mixing}, the moments of twist-3 distributions
receive contributions from the whole tower of  operators
with non-minimal anomalous dimensions at the level of $\CO(1/N_c^2)$
corrections. Aside from the ``energy denominators'',
the coefficients in front of these operators in \re{mixing}
are given by (normalized)
matrix elements $\vev{\ell'|{\cal H}^{(1)}|\ell=0}$ of the
perturbation ${\cal H}^{(1)}$
\begin{equation}
  \vev{\ell'|{\cal H}^{(1)}|\ell} =
  \frac{\vev{\Psi^{(0)}_{N,q_{\ell'}}|{\cal H}^{(1)}|\Psi^{(0)}_{N,q_\ell}}}
        {|\!|\Psi^{(0)}_{N,q_{\ell'}}|\!|\,|\!|\Psi^{(0)}_{N,q_\ell}|\!|},
\label{Hll'}
\end{equation}
where $\ell, \ell' = 0,1,\ldots, N$ numerate the anomalous
dimensions from below. Our main observation is that
the matrix elements (\ref{Hll'}) are
concentrated close to the diagonal $\ell=\ell'$, see Figs.~1,~2 and
Table~2.
\begin{figure}[ht]
\centerline{\epsfxsize9.0cm\epsffile{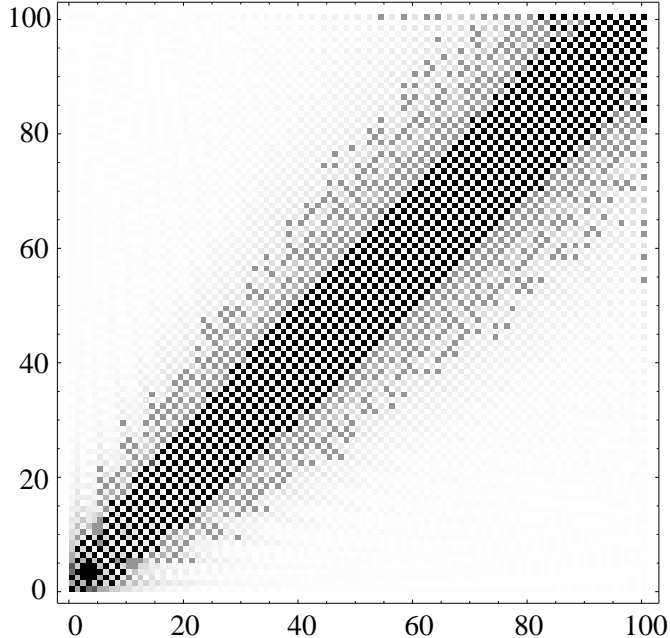}}
\caption[]{The density plot of the mixing matrix
$|\vev{\ell'|{\cal H}^{(1)}_{S^+}|\ell}|$
at $N=100$; $\ell=0$ corresponds to the lowest eigenstate
$\Psi^{S^+}_{N}$ of the large-$N_c$ Hamiltonian ${\cal
H}^{(0)}_{S^+}$. Notice ``chess-board'' structure
with alternating large (dark) and small (light) elements.}
\label{mixingS50}
\end{figure}
In addition, the
lowest anomalous dimensions $E_{N,q_0} \equiv E_N$ (\ref{E-ex})
are separated from the rest of the spectrum by a finite gap
$E_{N,q_1}-E_N = 0.227$ as $N\to\infty$ \cite{BDM98,DKM99}.
These two properties guarantee that the expansion in
\re{mixing} is rapidly converging and remains well defined
in the large-$N$ limit (when the number of eigenstates diverges).
The general structure for the cases of
${\cal H}_{S^+}$ and ${\cal H}_{T}$ is very similar; we present
the results for $S^+$ as related to the structure function
$g_2(x)$ and therefore being of more direct phenomenological
significance.

\begin{figure}[t]
\centerline{\epsfxsize10.0cm\epsffile{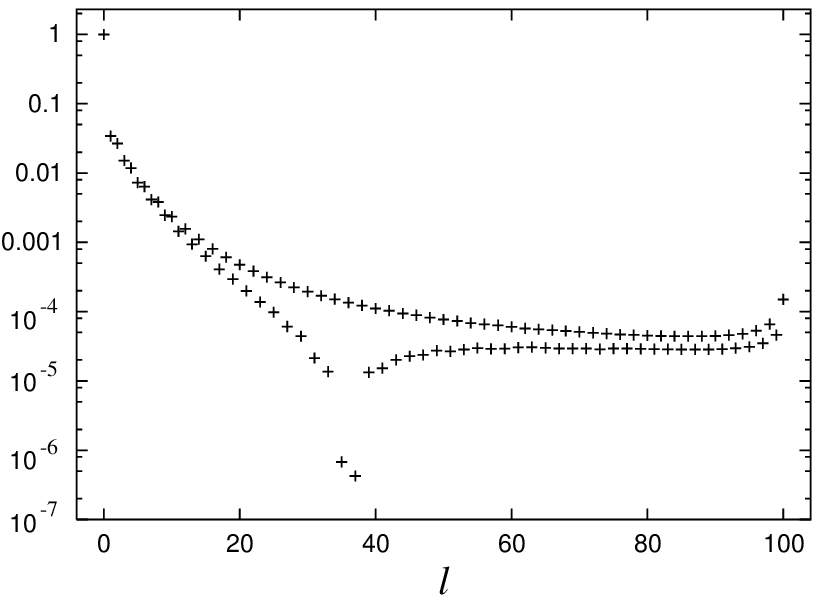}}
\caption[]{
Exact numerical results for the
coefficients $\frac{|\langle
\Psi_{N,q_\ell}|\Phi_{N}^{S}\rangle|}{
|\!|\Psi_{N,q_\ell}|\!| |\!|\Phi_{N}^{S}|\!|}$
that enter the expansion \re{decomp} of the
$N=100$th moment of $\widetilde g_2(x,Q^2)$ in
contributions of multiplicatively renormalized operators.
 Notice smallness of all coefficients with $\ell \ge
1$ and their different  behavior for even and odd $\ell$.
}
\label{slice100}
\end{figure}

\begin{table}[t]
\begin{center}
\begin{tabular}{|c||c|c|c|c|c|c|c|}
\hline
$\ell$ & $0$ &  $1$  & $2$ & $3$ & $4$ &$5$ &$6$ \\
\hline
${N=0}$
&$1.0000$ &        &        &        &        &  &\\
${N=1}$
&$0.9996$ & $0.0283$ &        &        &        &  &\\
${N=2}$
&$0.9991$ &$ 0.0080$ & $0.0412$ &        &        & & \\
${N=3}$
&$0.9995$ & $0.0037$ & $0.0295$ &$0.0063$ &        & &   \\
${N=4}$
&$0.9994$ &$ 0.0030$ & $0.0305$ &$0.0001$ & $0.0146$ & &    \\
${N=5}$
&$0.9995$ &$ 0.0056$ & $0.0278$ &$0.0005 $& $0.0096$ &$0.0037$ & \\
${N=6}$
&$0.9995$ & $0.0093$ &$ 0.0277$ &$0.0010$ &$ 0.0108$ &$0.0008$ &$
0.0078$\\
\hline
\end{tabular}
\end{center}
\caption{
Exact numerical results for the
coefficients $\frac{|\langle
\Psi_{N,q_\ell}|\Phi_{N}^{S}\rangle|}{
|\!|\Psi_{N,q_\ell}|\!| |\!|\Phi_{N}^{S}|\!|}$
that enter the expansion \re{decomp} of the lowest
moments of $\widetilde g_2(x,Q^2)$ in
contributions of multiplicatively renormalized operators.
Notice dominance of the coefficients with even $\ell$
over those with odd $\ell$ for a given $N$.
}
\label{Table2}
\end{table}

At large $N$
the matrix elements $\vev{\ell|{\cal H}^{(1)}|0}$ can be calculated
semi-classically using the approach developed in \cite{DKM99}.
It turns out that the matrix elements $|\vev{\ell|{\cal H}^{(1)}|0}|$
approach maximal value $\sim 1/\sqrt{\ln N}$
at $\ell \sim \ln N$, while for $\ell \gg \ln N$
they rapidly decrease as $1/\ell^2$.
As the result, the sum in~(\ref{mixing}) is effectively cut off at
$\ell \sim \ln N$ terms. For example,
for $N\sim 100$ only $\sim 5$ first (lowest) levels give a sizeable
contribution to the r.h.s.\ of \re{mixing}, cf. Fig.~2. Analytic
expressions for the
matrix elements $\vev{\ell|{\cal H}^{(1)}|0}$ in the large $N$
limit are complicated and will be given elsewhere \cite{2000}.

An independent argument in favor of the dominance of the operators
with lowest anomalous dimensions comes from the structure of matrix
elements $\langle\!\langle \CO_{N,q}\rangle\!\rangle$ in the large-$N$ limit.
In an analogy with twist-2 distributions, we can write the
matrix elements $\langle\!\langle \CO_{N,q}\rangle\!\rangle$ as
(generalized) moments of nonperturbative (chiral-odd or chiral-even)
distribution function $D_f(x_1,x_2,x_3)$ describing quark-gluon-antiquark
correlations in the nucleon, schematically%
\footnote{see Appendix A in \cite{DKM99} for
exact expressions.}
\begin{equation}
\langle\!\langle \CO_{N,q}\rangle\!\rangle=\int_{-1}^1 \! [\,dx\,]\,
\Psi_{N,q}(x_1,x_2,x_3)\,D_f(x_1,x_2,x_3),
\label{D}
\end{equation}
where $[dx]=dx_1\,dx_2\,dx_3\,\delta(x_1+x_2+x_3)$ and
the variables $x_i$ have the meaning of the quark, antiquark and gluon
momentum fractions, $-1\le x_i\le 1$.

At large $N$, the eigenfunctions $\Psi_{N,q}(x_i)$ are sharply peaked
at the boundary of the integration region $x_1+x_2+x_3=0$ so that
the matrix elements \re{D} probe the behavior of the
quark-gluon distribution functions $D(x_i)$ near the kinematical
boundaries. The precise position of the peak depends on $q$.
It turns out that the eigenfunctions of the
low-lying states are peaked at $x_1=-x_3=\pm 1,\,\,x_2=0$
corresponding to configurations when the quark and the antiquark carry
all the momentum of nucleon and the gluon is soft.
For  higher states the position of the peak  is shifted
gradually to $2x_1=2x_3=-x_2=\pm 1$ corresponding to a hard gluon and
(relatively) soft quark and antiquark. If one assumes that the
nonperturbative distributions are generated by gluon radiation
already at low scales, the configurations with soft gluons should be
enhanced and those with hard gluons suppressed.
We conclude that the contribution of the conformal operators
with large anomalous dimensions to the moments of the
structure functions, \re{mixing}, is suppressed because of smallness
of both {\it perturbative\/} mixing coefficients and
the corresponding {\it nonperturbative\/} matrix elements as describing
rare parton configurations in nucleon involving a hard gluon.

To summarize, we have argued that (up to $\CO(1/N_c^4)$ corrections)
the contribution of multiplicatively renormalizable operators with
higher anomalous dimensions to the moments of twist-3 structure functions
$e(x)$, $\widetilde h_L$ and $\widetilde g_2(x)$ is small as compared with
that of the ground state operator.
With the increasing precision of the experimental data
this contribution can be estimated in a ``two-channel''
approximation
in which one supplements the dominant ground state quark-gluon component with
one extra effective partonic component
in order
to account for the admixture of $\sim \ln N$ lowest eigenstates.
\vskip0.3cm

{\large\bf Acknowledgements.} Work by A.M. was partially
supported by the DFG.

\end{document}